\documentclass[10pt,a4paper]{article}
\usepackage[dvips]{graphicx}
\usepackage{latexsym}
\usepackage{amsmath,amssymb}

\title{Non-Minimal Two-Loop Inflation}
\author{Tomohiro Inagaki${}^{a}$, Ryota Nakanishi${}^{b}$, and Sergei D. Odintsov${}^{c,d,e,f}$,\\
${}^{a}$Information Media Center, Hiroshima University, Higashi-Hiroshima, 739-8521, Japan,\\
${}^{b}$Department of Physics, Hiroshima University, Higashi-Hiroshima, 739-8526, Japan,\\
${}^{c}$Instituto de Ciencias del Espacio (ICE/CSIC) and Institut de Estudis Espacials\\ de Catalunya (IEEC), Campus UAB,
Carrer de Can Magrans, s/n 08193 \\Cerdanyola del Vall\'{e}s (Barcelona), Spain \\
${}^{d}$Instituci\'{o} Catalana de Recerca i Estudis Avan\c{c}ats (ICREA), Barcelona, Spain\\
${}^{e}$National Research Tomsk State University, 634050 Tomsk and Tomsk State \\Pedagogical University, 634061 Tomsk, Russia\\
${}^{f}$Inst.of Physics, Kazan Federal University, 420008 Kazan,Russia
}

\usepackage[top=30truemm,bottom=30truemm,left=25truemm,right=25truemm]{geometry}

\makeatletter
\@addtoreset{equation}{section}
\makeatother

\begin{document}
 \maketitle
 \begin{abstract}

We investigate the chaotic inflationary model using the two-loop effective potential of a self-interacting scalar field theory in curved spacetime. We use the potential  which contains a non-minimal scalar curvature coupling and a quartic scalar self-interaction and which was found in Ref. \cite{Odintsov:1993rt}. We analyze the Lyapunov stability of de Sitter solution and show the stability bound. Calculating the inflationary parameters, we systematically explore the spectral index $n_s$ and the tensor-to-scalar ratio $r$, with varying the four parameters, the scalar-curvature coupling $\xi_0$, the scalar quartic coupling $\lambda_0$, the renormalization scale $\mu$ and the e-folding number $N$. It is found that the two-loop correction on $n_s$ is much larger than the leading-log correction, which has previously been studied in Ref. \cite{Inagaki:2014wva}. We show that the model is consistent with the observation by Planck with WMAP \cite{Hinshaw:2012aka,Ade:2013uln} and a recent joint analysis of BICEP2 \cite{Ade:2015tva}.

\noindent
 {\it keyword:} two-loop effective potential, spectral index, tensor-to-scalar ratio 
 
 \end{abstract} 

 \section{Introduction}
The measurements of the cosmic microwave background (CMB) fluctuations become increasingly important from the perspective of not only cosmology but also elementary particle physics. Useful indicators of the CMB fluctuations are given by the scalar (or density) fluctuations, $\delta$, the spectral index, $n_s$, and the tensor-to-scalar ratio, $r$. In the inflation scenario a non-vanishing potential energy density of a scalar field induces exponential expansion on the universe. The origin of the CMB fluctuations is found in the quantum fluctuations of the scalar field.  It is expected that these inflationary parameters restrict the models of particle physics. Although we have observed only one elementary scalar field, i.e. Higgs, it is quite natural to assume that other scalar fields exist and play a decisive role for the energy density and its fluctuation at early universe.

In this paper we consider that the inflaton field is a real scalar field with a quartic self-interaction and a non-minimal scalar-curvature interaction at high-energy scale and study a possible model consistent with the CMB fluctuations at the two loop level. Inflation is thought to occur near the Planck scale. In such high energy scale the quantum correction may have some remarkable effect on the inflationary parameters. It is known that the spectral index and the tensor-to-scalar ratio are independent of the scalar quartic coupling, $\lambda_0$, at the tree level. It is also known that there is an attractor on the $(n_s, r)$ plane. The inflationary parameters, $n_s$ and $r$, converge to their universal model-independent values at the large scalar-curvature coupling limit \cite{Kallosh:2013maa,Kallosh:2013tua}. Note that RG behavior of scalar curvature coupling $\xi$ is defined by the behavior of the corresponding quantum field theory at high energy (see: \cite{Muta:1991mw,Buchbinder}) so that it maybe tend to large or small asymptotic value at high-energy limit.

The inflationary parameters have been investigated up to the leading log level with respect to the scalar quartic coupling in Ref. \cite{Inagaki:2014wva}. The quantum corrections introduce the quartic coupling dependence for $n_s$ and $r$, but do not alter the attractor behavior. The standard model (SM) Higgs inflation has been investigated up to the next to leading log level in Refs. \cite{Bezrukov:2009db, Allison:2013uaa}. In SM the scalar quartic coupling is extremely suppressed near the Planck scale. It has been pointed out that a large non-minimal scalar curvature coupling is necessary to reproduce the observed Higgs mass, $n_s$ and $r$. The remark is in order. Quantum field theory in curved spacetime induces log terms in the scalar four-point as well as in the non-minimal 
scalar-curvature sector, for the corresponding effective potential \cite{Elizalde:1993ee,Elizalde:1993qh}. The account of such quantum-corrected terms in the potential, especially RG improved effective potential, is done for the study of inflation in \cite{Buchel:2004df,DeSimone:2008ei,Lee:2013nv,Okada:2013vxa,Barenboim:2013wra,Hamada:2014iga,Oda:2014rpa,Ren:2014sya,Hamada:2014xka,Elizalde:2014xva,Hamada:2014wna,He:2014ora,Herranen:2014cua}. It is also interesting to note that the reconstruction of the inflationary scalar potential as is done in \cite{Bamba:2014daa} maybe applied to such non-minimal inflationary scalar potential.

The paper is organized as follows. Following Ref.\cite{Odintsov:1993rt} we introduce the effective Lagrangian up to the two-loop level in Sec.~2. The Lagrangian in the Einstein frame is formulated by the Weyl transformation. We show the concrete expressions for the scalar fluctuations, the spectral index and the tensor-to-scalar ratio at the two loop level in Sec.3. The Lyapunov stability of de Sitter solutions is studied in Sec.~4 in close analogy with the corresponding study for log-corrected higher-derivative quantum gravity \cite{Bamba:2014mua}. In Sec.5, we numerically evaluate the inflationary parameters and show possible parameters consistent with the observed data. Finally we give some concluding remarks.

 \section{Two-loop effective Lagrangian in Einstein frame}
Here we consider a massless scalar field with a non-minimal scalar-curvature coupling, $\xi_0$, and adopt
a simple chaotic inflation scenario near the Planck scale. We start from a Lagrangian density,
\begin{eqnarray}
\mathcal{L}^{(J)} &=&  \sqrt{-g}\left(\frac{1}{2}R + \frac{1}{2}\xi_0 R\phi^2 - \frac{1}{2}g^{\mu\nu} (\partial_{\mu} \phi)( \partial_{\nu}\phi) - V^{(J)}\right), \label{lag:J} \\
V^{(J)} &=& \frac{\lambda_0}{24}\phi^4,
\end{eqnarray}
where the superscript $(J)$ denotes the Jordan frame, $g$ is the determinant of the metric tensor, $g_{\mu\nu}$, $\xi_0$, and $\lambda_0$ represents the scalar-curvature and scalar quartic couplings, respectively. The Jordan frame is characterized by the existence of the $\xi$-term. Here the reduced Planck mass is set as $M_{p}=(8\pi G)^{-1/2}=1$.

In Ref. \cite{Odintsov:1993rt} the closed expression for the two-loop effective potential is given under the linear curvature approximation,
\begin{eqnarray}
V &=& \frac{\lambda_0}{24}\phi^4 - \frac{1}{2}\xi_0R\phi^2 + \frac{\lambda_0^2\phi^4}{(16\pi)^2}\ln{\frac{\phi^2}{\mu^2}} - \frac{\lambda_0(\xi_0-1/6)}{(8\pi)^2}R\phi^2\ln{\frac{\phi^2}{\mu^2}} - \frac{\lambda_0^3\phi^4}{8(4\pi)^4}\ln{\frac{\phi^2}{\mu^2}+ \frac{3\lambda_0^3\phi^4}{32(4\pi)^4}\left(\ln{\frac{\phi^2}{\mu^2}}\right)^2 } \nonumber \\
&\ &\ \ \ -\frac{\lambda_0^2}{4(4\pi)^4}\left[\left(\xi_0-\frac{1}{6}\right)+\frac{1}{36}\right]R\phi^2\ln{\frac{\phi^2}{\mu^2}} - \frac{\lambda_0^2(\xi_0-1/6)}{4(4\pi)^4}R\phi^2\left(\ln{\frac{\phi^2}{\mu^2}}\right)^2,
\label{pot}
\end{eqnarray}
where $\mu$ represents the renormalization scale. It should be noted that some terms in the effective potential vanish if we set the scalar-curvature coupling as $\xi_0=1/6$, called the conformal one. 
From Eqs.(\ref{lag:J})-(\ref{pot}) we define effective scalar-dependent couplings, $\xi$ and $\lambda$, as
\begin{eqnarray}
\frac{1}{2}\xi &\equiv& \frac{1}{2}\xi_0 + \frac{\lambda_0(\xi_0 - 1/6)}{(8\pi)^2}\ln{\frac{\phi^2}{\mu^2}} \nonumber \\ && +\frac{\lambda_0^2}{4(4\pi)^4}\left[\left(\xi_0-\frac{1}{6}\right)+\frac{1}{36}\right]\ln{\frac{\phi^2}{\mu^2}} + \frac{\lambda_0^2(\xi_0-1/6)}{4(4\pi)^4}\left(\ln{\frac{\phi^2}{\mu^2}}\right)^2, \label{rep1} \\
\frac{\lambda}{24} &\equiv& \frac{\lambda_0}{24} + \frac{\lambda_0^2}{(16\pi)^2}\ln{\frac{\phi^2}{\mu^2}} - \frac{\lambda_0^3}{8(4\pi)^4}\ln{\frac{\phi^2}{\mu^2}} + \frac{3\lambda_0^3}{32(4\pi)^4}\left(\ln{\frac{\phi^2}{\mu^2}}\right)^2. \label{rep2}
\end{eqnarray}
Thus the two-loop effective Lagrangian is given by
\begin{eqnarray}
\mathcal{L}_{eff}^{(J)} &=& \sqrt{-g}\left(\frac{1}{2}R + \frac{1}{2}\xi R\phi^2 - \frac{1}{2}g^{\mu\nu} (\partial_{\mu} \phi)( \partial_{\nu}\phi) - V^{(J)}_{eff} \right), \label{lag:eff:J} \\
V^{(J)}_{eff} &=& \frac{\lambda}{24}\phi^4.
\label{pot:eff:J}
\end{eqnarray}
The Lagrangian, $\mathcal{L}_{eff}^{(J)}$, has a similar form with (\ref{lag:J}) thanks to the definitions (\ref{rep1}) and (\ref{rep2}). 
This potential has the renormalization scale dependence which stems from the radiative corrections for $\phi^4$ theory in curved spacetime.

For calculations of the inflationary parameters, it is more convenient to change the frame into the Einstein frame where the $\xi$-term disappears. In order to change the frame, we consider the Weyl transformation,
\begin{equation}
\tilde{g}^{\mu\nu} = \Omega^{-2}(x) g^{\mu\nu},
\label{Weyl}
\end{equation}
where $\tilde{g}^{\mu\nu}$ is the metric tensor in the transformed frame. The Weyl factor, $\Omega$, is an analytic function with respect to the space-time coordinate.
 After this Weyl transformation the two-loop effective Lagrangian is rewritten as
\begin{eqnarray}
\mathcal{L}^{(J)}_{eff} &\rightarrow& \Omega^{-2}\sqrt{-\tilde{g}}\biggl[\frac{1}{2}(1+\xi\phi^2)\tilde{R} - \frac{1}{2}\tilde{g}^{\mu\nu}(\partial_{\mu} \phi) (\partial_{\nu} \phi) - \Omega^{-2}V^{(J)}_{eff} \nonumber \\
&\ &\ \ \ \ \ \ \ \ \ \ \ \ \ \ \ \ \ \ \ \ + 3\left[\tilde{\square}\ln{\Omega} - \tilde{g}^{\mu\nu}(\partial_{\mu}\ln{\Omega})(\partial_{\nu} \ln{\Omega})\right]\left(1 + \xi\phi^2 \right)\biggr],
\end{eqnarray}
where $\tilde{g}$, $\tilde{R}$ and $\tilde{\square}$ are the determinant of the metric tensor, the Ricci scalar and the d'Alembert operator in the transformed frame, respectively. We can transform the Jordan frame into the Einstein frame by choosing the Weyl factor in order that the $\xi$-term is eliminated. The suitable choice is
\begin{equation}
\Omega^2 = 1 + \xi \phi^2.
\label{para:Weyl}
\end{equation}
Then we redefine the scalar field to obtain the canonical kinetic term for the scalar field. The redefined scalar field is given by the relation,
\begin{equation}
\frac{\partial\varphi}{\partial\phi} = \frac{\sqrt{\Omega^2 + \cfrac{3}{2} \left(\cfrac{\partial\Omega^2}{\partial\phi}\right)^2}}{\Omega^2},
\label{FR}
\end{equation}
where $\varphi$ is the redefined canonical scalar field. With these techniques we finally obtain the Lagrangian in the Einstein frame,
\begin{eqnarray}
\mathcal{L}^{(E)}_{eff} &=& \sqrt{-\tilde{g}}\left[\frac{1}{2}\tilde{R} - \frac{1}{2}\tilde{g}^{\mu\nu} \partial_{\mu} \varphi \partial_{\nu} \varphi - V^{(E)}_{eff} \right], \label{lag:E} \\
V^{(E)}_{eff} &=& \Omega^{-4}V^{(J)}_{eff},
\label{pot:E}
\end{eqnarray}
where the superscript $(E)$ represents the Einstein frame. 
The effective potential (\ref{pot:E}) has the $\xi$- and $\mu$-dependences in addition to the $\lambda$-dependence. The first one comes from the Weyl transformation and the second one from the quantum corrections.

 \section{Inflationary parameters}
CMB fluctuations arise from the quantum fluctuations of the inflaton field. The measurements of CMB fluctuations directly restrict the inflationary parameters. Within the context of the slow-roll scenario, the inflationary parameters are fully represented by means of the inflaton potential. The e-folding number, $N$, the slow-roll parameters, $\epsilon$ and $\eta$, are formulated as \cite{Kallosh:2013tua},
\begin{eqnarray}
N &=& \int_{\phi_{end}}^{\phi_N} \left(\frac{\partial\varphi}{\partial\phi}\right)^2 \frac{V^{(E)}}{\partial V^{(E)} /\partial\phi}\ d\phi, 
\label{cp1}\\
\epsilon &=& \frac{1}{2}\left(\frac{1}{V^{(E)}}\frac{\partial V^{(E)}}{\partial\phi}\frac{\partial\phi}{\partial\varphi} \right)^2, 
\label{cp2}\\
\eta &=& \frac{1}{V^{(E)}}\left[\frac{\partial}{\partial\phi}\left(\frac{\partial V^{(E)}}{\partial\phi}\frac{\partial\phi}{\partial\varphi}\right) \right]\frac{\partial\phi}{\partial\varphi}, 
\label{cp3}
\end{eqnarray}
where the upper and lower limits of the integral, $\phi_N$ and $\phi_{end}$, are the field configurations when the slow-rolls scenario starts and brakes respectively. It is assumed that the inflaton slowly rolls downhill of the potential from $V(\phi_N)$ to $V(\phi_{end})$. 
The formula of scalar (or density) fluctuations $\delta$ is given in Ref.~\cite{Linde:2005ht}. We modify it including the field redefinition (\ref{FR}),
\begin{equation}
\delta = \left. \frac{\bigl(V^{(E)}\bigr)^{3/2}}{\sqrt{12\pi^2}} \Biggl( \frac{\partial V^{(E)}}{\partial \phi} \cfrac{\partial \phi}{\partial \varphi} \Biggr)^{-1} \right|_{\phi=\phi_N},\label{cp3.5}
\end{equation}
where the constant, $C$, appearing in Ref.~\cite{Linde:2005ht} is taken to be $1$.

We apply these formulae to the two-loop effective potential.
Substituting the effective potential (\ref{pot:E}) into Eqs. (\ref{cp1})-(\ref{cp3.5}), the inflationary parameters in the two-loop $\phi^4$ model read
\begin{eqnarray}
N &=& \int_{\phi_{end}^2}^{\phi_N^2} \left(\frac{\partial\varphi}{\partial\phi}\right)^2
\frac{1}{2\phi(\ln V^{(E)})' } d\phi^2,
\label{cp4}\\
\epsilon &=& \cfrac{1}{2}\left[ (\ln V^{(E)})'  \right]^2 \left(\frac{\partial\phi}{\partial\varphi}\right)^2,
\label{cp5} \\
\eta &=& \Biggl[ -\frac{16}{\phi^2} -\frac{2(\Omega^{2'})^2}{\Omega^4} - \frac{2\Omega^{2''}}{\Omega^2} +\frac{14\Omega^{2'}}{\Omega^2\phi} + \frac{18\lambda_0^3}{(4\pi)^4\lambda\phi^2}+7\frac{(\ln V^{(E)})' }{\phi} -\frac{4\Omega^{2'}(\ln V^{(E)})' }{\Omega^2}
  \nonumber \\
&\ \ \ &\ \  +  (\ln V^{(E)})' \frac{3\Omega^2\Omega^{2'} + 3(\Omega^{2'})^3 - 3\Omega^2\Omega^{2'}\Omega^{2''}}{2\Omega^4}\left(\frac{\partial\phi}{\partial\varphi}\right)^2
\Biggr]\left(\frac{\partial\phi}{\partial\varphi}\right)^2, \label{cp6}\\
\delta &=& \left. \frac{1}{12\pi (\ln V^{(E)})' }\sqrt{\cfrac{\lambda}{2}} \phi^2\cfrac{\partial\varphi}{\partial\phi}\right|_{\phi=\phi_N}, \label{cp7}
\end{eqnarray}
with
\begin{eqnarray}
(\ln V^{(E)})' &\equiv& \frac{1}{V^{(E)}}\frac{\partial V^{(E)}}{\partial\phi} =\cfrac{4}{\phi} + \cfrac{6}{\lambda\phi\ln{\phi^2/\mu^2}}\left[ \cfrac{\lambda-\lambda_0}{3} + \cfrac{3\lambda_0^3}{4(4\pi)^4}\left(\ln{\cfrac{\phi^2}{\mu^2}}\right)^2 \right]  -2\cfrac{\Omega^{2'}}{\Omega^2},\\
\Omega^{2'} &\equiv& \frac{\partial\Omega^2}{\partial\phi} 
\ =\  2\xi\phi + \frac{2\phi}{\ln{\phi^2/\mu^2}}\left[ \xi - \xi_0 +  \frac{\lambda_0^2(\xi_0-1/6)}{2(4\pi)^4}\left(\ln{\frac{\phi^2}{\mu^2}}\right)^2 \right], \\ 
\Omega^{2''} &\equiv& \frac{\partial^2\Omega^2}{\partial\phi^2} 
\ =\   2\xi + \frac{2}{\ln{\phi^2/\mu^2}}\left[ 3(\xi-\xi_0) + \frac{\lambda_0^2(\xi_0-1/6)}{2(4\pi)^4}\ln{\frac{\phi^2}{\mu^2}}\left(4 + 3\ln{\frac{\phi^2}{\mu^2}}\right)  \right].
\label{Omega2}
\end{eqnarray}
Under the slow-roll approximation the spectral index, $n_s$, and the tensor-to-scalar ratio, $r$, are given by
\begin{eqnarray}
n_s &=& 1 + 2\left.\eta\right|_{\phi=\phi_N} - 6\left. \epsilon\right|_{\phi=\phi_N}  ,
\label{ns}\\
r &=& 16\left. \epsilon\right|_{\phi=\phi_N}  .
\label{r}
\end{eqnarray}
These expressions are valid only if the slow-roll parameters are small enough, $\epsilon \ll 1$ and $|\eta|\ll 1$. The field configuration $\phi_{end}$ is fixed by the boundary condition where the slow-roll parameters exceed the value of order one. 

It is instructive to take the small coupling limit, $\lambda_0\to 0$. At this limit we can analytically carry out the integration in Eq (\ref{cp4}) and obtain the simpler forms for the inflationary parameters (\ref{cp4})-(\ref{cp7}),
\begin{eqnarray}
N &=& \frac{1}{8} \left[ (1+6\xi_0)(\phi_N^2 - \phi_{end}^2) -6\ln{\frac{1+\xi_0 \phi_N^2}{1+\xi_0 \phi_{end}^2}} \right],
\label{lim:N:nolam}\\
\epsilon &=& \frac{8}{\phi^2(1 +\xi_0 \phi^2 + 6\xi_0^2 \phi^2)},
\label{lim:ep:nolam}\\
\eta &=& \frac{4(3 + \xi_0 \phi^2 +12 \xi_0^2 \phi^2 - 2\xi_0^2 \phi^4 - 12\xi_0^3 \phi^4)}{\phi^2(1+\xi_0 \phi^2 + 6\xi_0^2 \phi^2)^2},
\label{lim:et:nolam}\\
\delta &=& \frac{\sqrt{\lambda_0}\phi_{N}^3}{48\sqrt{2}\pi} \frac{\sqrt{1+\xi_0 \phi_{N}^2 +6\xi_0^2 \phi_{N}^2}}{1+\xi_0 \phi_{N}^2}.
\label{lim:de:nolam}
\end{eqnarray}
The equations (\ref{lim:N:nolam})-(\ref{lim:et:nolam}) coincide with the expressions in Ref.~\cite{Kallosh:2013maa}. Apart from the scalar fluctuations (\ref{lim:de:nolam}), the inflationary parameters are found to be independent of $\lambda_0$ at the limit, $\lambda_0\rightarrow 0$. Thus the spectral index, $n_s$, and the tensor-to-scalar ratio, $r$, are determined by setting the scalar-curvature coupling, $\xi_0$. 
For $\phi_N\gg \phi_{end}$, the slow-roll parameters reduce to
\begin{equation}
\left.\epsilon\right|_{\phi=\phi_N} \rightarrow \frac{3}{4N^2},\ \ \ 
\left.\eta\right|_{\phi=\phi_N} \rightarrow -\frac{1}{N}
\label{lim:strong:epet}
\end{equation}
at the large $\xi_0$ limit. Thus the inflationary parameters, $r$ and $n_s$, approach to
\begin{equation}
r \rightarrow \frac{12}{N^2},\ \ \ 
n_s \rightarrow 1-\frac{2}{N}.
\label{lim:strong:rns}
\end{equation}
This is just the universal attractor mentioned in Ref.~\cite{Kallosh:2013tua}.
On the other hand, the slow-roll parameters are also simplified at the small $\xi_0$ limit,
\begin{equation}
\left.\epsilon\right|_{\phi=\phi_N} \rightarrow \frac{1}{N},\ \ \ 
\left.\eta\right|_{\phi=\phi_N} \rightarrow \frac{3}{2N}.
\label{lim:weak:epet}
\end{equation}
At this limit the asymptotic values of $r$ and $n_s$ are given by
\begin{equation}
r \rightarrow \frac{16}{N},\ \ \ 
n_s \rightarrow 1-\frac{3}{N}.
\label{lim:weak:rns}
\end{equation}
As is pointed out in Ref.~\cite{Kallosh:2014laa}, Eq. (\ref{lim:weak:rns}) does not represent the attractor. It is the model-dependent result in the various chaotic inflation models with the minimal curvature coupling.

 \section{Lyapunov stability}
The spacetime of the inflationary universe is thought to be approximately de Sitter space. The exponential evolution of the spacetime needs to become unstable in order to get exit to radiation/matter dominant eras. Thus to check the stability of the de Sitter solution is important. If we suppose the spatially flat Friedmann-Lemaitre-Robertson-Walker (FLRW) universe during the inflationary era, Lyapunov stability of the de Sitter solution can be evaluated by the following formula \cite{ Elizalde:2014xva, Lyapunov, Pontryagin},
\begin{equation}
K_f \equiv \left. \frac{V'U'+2VU''-UV''}{3(U')^{2}+U}\right|_{\phi=\phi_f}, \label{Kf}
\end{equation}
where the function $V(\phi)$ is the inflaton potential and $U(\phi)$ is the term multiplied by the curvature $R$ in the Lagrangian in Jordan frame. Notice that, in this paper, the Lyapunov stability $K_f$ is evaluated in Jordan frame. The function $V'$ and $V''$ denote the first and second derivatives of $V$ with respect to $\phi$. The de Sitter solution is unstable for a positive $K_f $ and stable for a negative $K_f$. The fixed point, $\phi_f$, is determined by
\begin{equation}
\left. 2\frac{U'}{U}\right|_{\phi=\phi_f} = \left. \frac{V'}{V}\right|_{\phi=\phi_f} .
\label{condition}
\end{equation}

\begin{figure}
  \begin{center}
   \includegraphics[width=71mm]{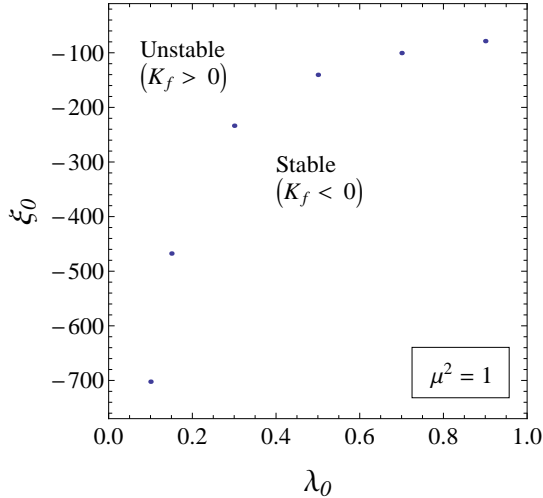}
  \end{center}
    \caption{The phase structure for the Lyapunov stability for $\mu^2=1$.}
\label{fig:stability}
\end{figure}

Substituting the Weyl factor (\ref{para:Weyl}) to $U$ and the effective potential (\ref{pot:eff:J}) to $V$, we calculate $K_f$ for the model (\ref{lag:eff:J}). There is no solution of Eq. (\ref{condition})  for a positive scalar-curvature coupling $\xi_0$.
For a negative $\xi_0$ some field configurations satisfy Eq. (\ref{condition}). The phase structure of Lyapunov stability for a negative $\xi_0$ is illustrated in Fig. \ref{fig:stability}.
It is found that the de Sitter solution is unstable when the absolute value of the scalar-curvature coupling, $\xi_0$, or the quartic scalar coupling, $\lambda_0$, is small enough. The figure \ref{fig:stability} is plotted for $\mu^2=1$. The renormalization scale dependence is very small. The difference in the results is only a few percent for a wide range of scales from $\mu^2=10^{-6}$ to $10^3$.

It should be noted that the exit from an approximately de Sitter space can be achieved through the reheating process. At this era we should carefully consider the interactions of the inflaton to light particles which assure the reheating of the universe \cite{NeferSenoguz:2008nn}.
The field configuration, $\phi_f$, should have a value between $\phi_N$ and $\phi_{end}$ as long as one assumes the slow-roll inflation scenario. In the considered model we cannot get the appropriate values of $\phi_N$ and $\phi_{end}$ for a negative $\xi_0$. It is unidentified whether the solution, $\phi_f$, lies in the inflationary era or not.
However, we believe that the above result shows something true and the de Sitter solution is unstable for a positive $\xi_0$. In the next section we consider only positive values of the scalar-curvature coupling, $\xi_0$, and numerically evaluate the inflationary parameters.

 \section{Numerical results}
In Sec.3 we have analytically derived the explicit expressions for the inflationary parameters. These parameters depend on the scalar-curvature coupling, $\xi_0$, the scalar quartic coupling, $\lambda_0$, and the renormalization scale, $\mu$. These parameters also get the dependence of the e-folding number $N$ through $\phi_N$. In this section we numerically investigate the behavior of the spectral index, $n_s$, and the tensor-to-scalar ratio, $r$, with varying the four model parameters, $\xi_0$, $\lambda_0$, $\mu^2$ and $N$. The results are compared with the ones at the leading-log level obtained in Ref.~\cite{Inagaki:2014wva}.

\begin{figure}
  \begin{center}
   \includegraphics[width=72mm]{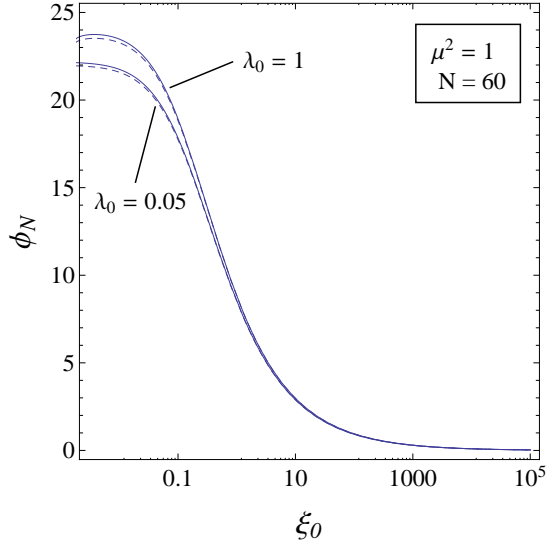}
  \end{center}
  \caption{Behavior of $\phi_N$ as a function of $\xi_0$ for $N=60$ and $\mu^2=1$. The solid and dashed lines are drawn for the two-loop and leading-log results, respectively.}
\label{fig:phiN}
\end{figure}

First we determine the field configurations at the beginning and the end of the slow-roll scenario, $\phi_{N}$ and $\phi_{end}$. At the end of the scenario the orders of the slow-roll parameters exceed unity. Thus the end of the scenario can be determined by
\begin{equation}
\left.\epsilon\right|_{\phi=\phi_{end}}=1 .
\end{equation}
Numerically performing the integration in Eq.~(\ref{cp4}) with $N$ fixed, we obtain $\phi_N$. The inflationary parameters should be evaluated at the horizon crossing, $\phi=\phi_N$, and so get the $N$ dependence. In Fig.~\ref{fig:phiN} we show the behavior of the field configuration, $\phi_N$, as a function of the scalar-curvature coupling, $\xi_0$. The field configuration, $\phi_N$, strongly depends on $\xi_0$, and is much greater than $\phi_{end}$, as seen in Fig.~\ref{fig:phiN}. The solid and dashed lines represent the results for the two-loop potential and leading-log potential, respectively. The difference between the results from the leading-log potential and the two-loop potential is very small, less than $2\%$.

\begin{figure}
  \begin{center}
   \includegraphics[width=68mm]{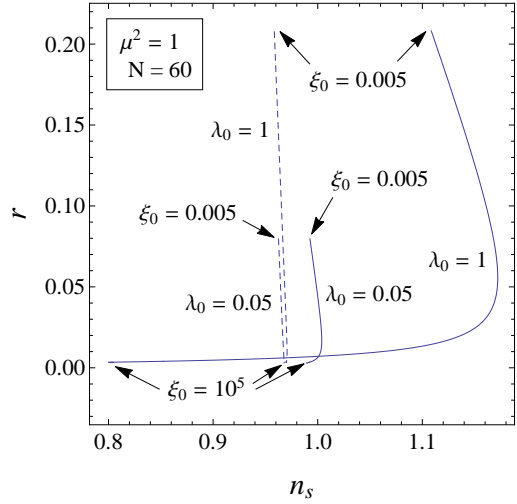}
  \end{center}
  \caption{Behavior of $n_s$ and $r$ as a function of the scalar-curvature coupling $\xi_0$ in the interval from $0.005$ to $10^5$ for $N=60$, $\mu^2=1$ and $\lambda_0 = 0.05,1$. The solid and dashed lines are drawn for the two-loop and leading-log results respectively.}
\label{fig:nsr:xi0}
\end{figure}

The slow-roll parameters, $\left.\epsilon\right|_{\phi=\phi_{N}}$ and $\left.\eta\right|_{\phi=\phi_{N}}$, are numerically calculated by Eqs.~(\ref{cp5}) and (\ref{cp6}). Then we obtain the spectral index, $n_s$, and tensor-to-scalar ratio, $r$ from Eqs.~(\ref{ns}) and (\ref{r}).  In Fig.~\ref{fig:nsr:xi0} we illustrate the $\xi_0$-dependence of $n_s$ and $r$ for $0.005<\xi_0<10^5$. It is observed that the two-loop correction to $n_s$ is larger than the leading-log correction. The difference increases for a larger $\lambda_0$. It is found that, whether for the leading-log or for the two-loop, the tensor-to-scalar ratio approaches $r\simeq0.003$ independent of $\lambda_0$, as $\xi_0$ increases. We cannot reproduce the attractor behaviors in \cite{Kallosh:2013maa,Kallosh:2013tua} in the range, $0.005<\xi_0<10^5$. However it is not adequate to deny the existence of the attractors out of this range, i.e. a larger $\xi_0 \gg10^5$.

\begin{figure}
 \begin{minipage}{0.48\hsize}
  \begin{center}
   \includegraphics[width=72mm]{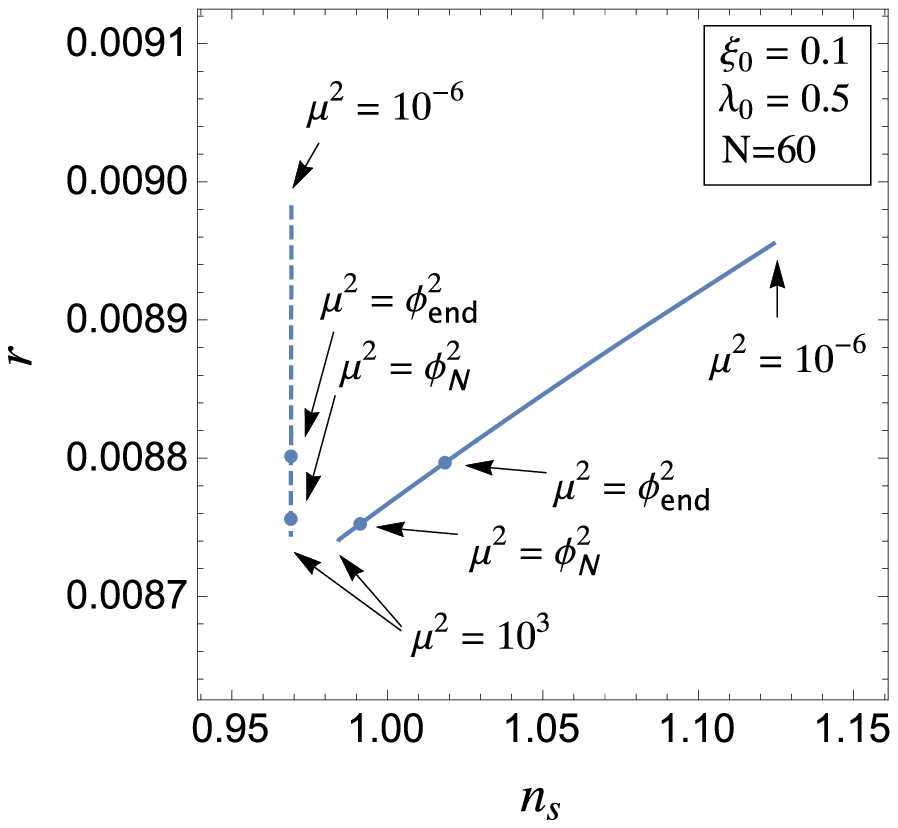}
  \end{center}
 \end{minipage}
 \hspace{0.04\hsize}
 \begin{minipage}{0.48\hsize}
  \begin{center}
   \includegraphics[width=72mm]{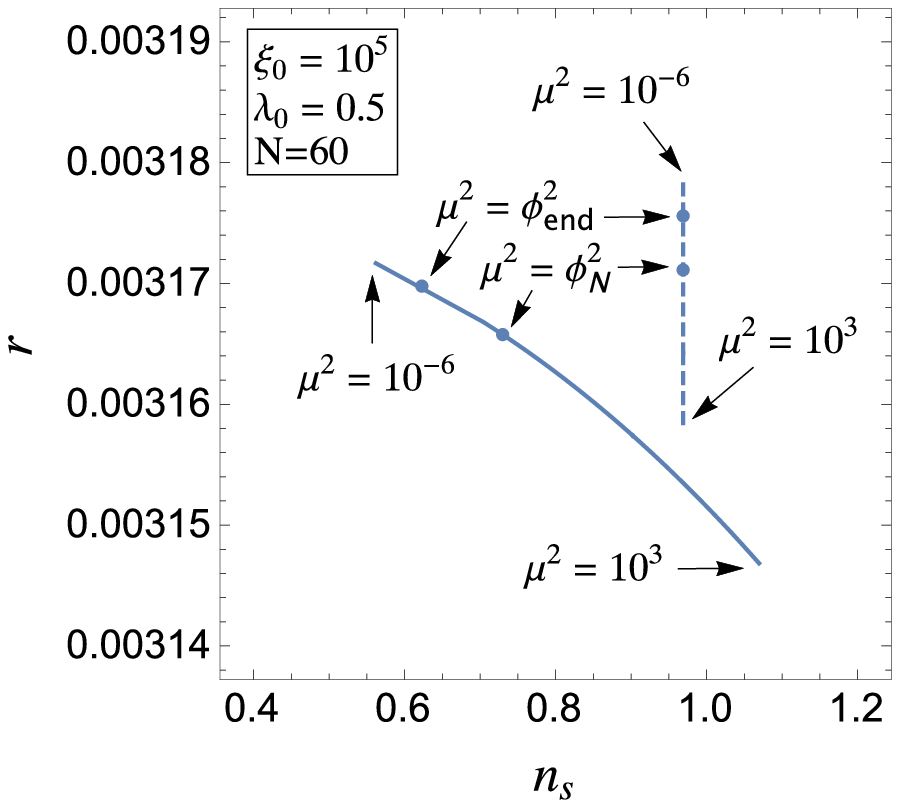}
  \end{center}
 \end{minipage}
    \caption{Behavior of $n_s$ and $r$ as a function of the renormalization scale, $\mu^2$, in the interval from $10^{-6}$ to $10^3$ for $N=60$, $\lambda_0=0.5$ and $\xi_0 = 0.1,10^5$. The solid and dashed lines are drawn for the two-loop and leading-log results respectively.}
\label{fig:nsr:mu2}
\end{figure}

The renormalization scale dependence is illustrated in Fig.\ref{fig:nsr:mu2}. We note that the lower value, $\mu^2=10^{-6}$, corresponds to a typical GUT scale. It is observed that the tensor-to-scalar ratio, $r$, decreases as $\mu^2$ increases. This feature for the renormalization scale dependence is similar to that for $\xi_0$ in both two-loop and leading-log results. For $\xi_0=0.1$ the tensor-to-scalar ratio takes the minimum value, $r\simeq0.00874$, at $\mu^2=10^3$ in both results. It is worth noting that for a weak scalar-curvature coupling, $\xi_0=0.1$, $n_s$ increases as $\mu^2$ decreases in the two-loop result, while it is almost fixed for $10^{-6}<\mu^2<10^3$ in the leading-log result. An opposite $\mu^2$-dependence of $n_s$ is observed for $\xi_0=10^5$ in the two-loop result, unlike the leading-log result. 
It is considered that there is an optimized scale which minimizes the quantum corrections at a certain value of the inflaton field \cite{Bezrukov:2009db}. It is normally set at the horizon exit, $\mu^2=\phi_N^2$, or at the end of the slow-roll scenario, $\mu^2=\phi_N^2$ \cite{Enqvist:2013eua}. These optimized scales are also plotted in Fig.\ref{fig:nsr:mu2}. A smaller two-loop contribution is observed for the renormalization scale, $\mu^2=\phi_N^2$.

\begin{figure}
 \begin{minipage}{0.48\hsize}
  \begin{center}
   \includegraphics[width=68mm]{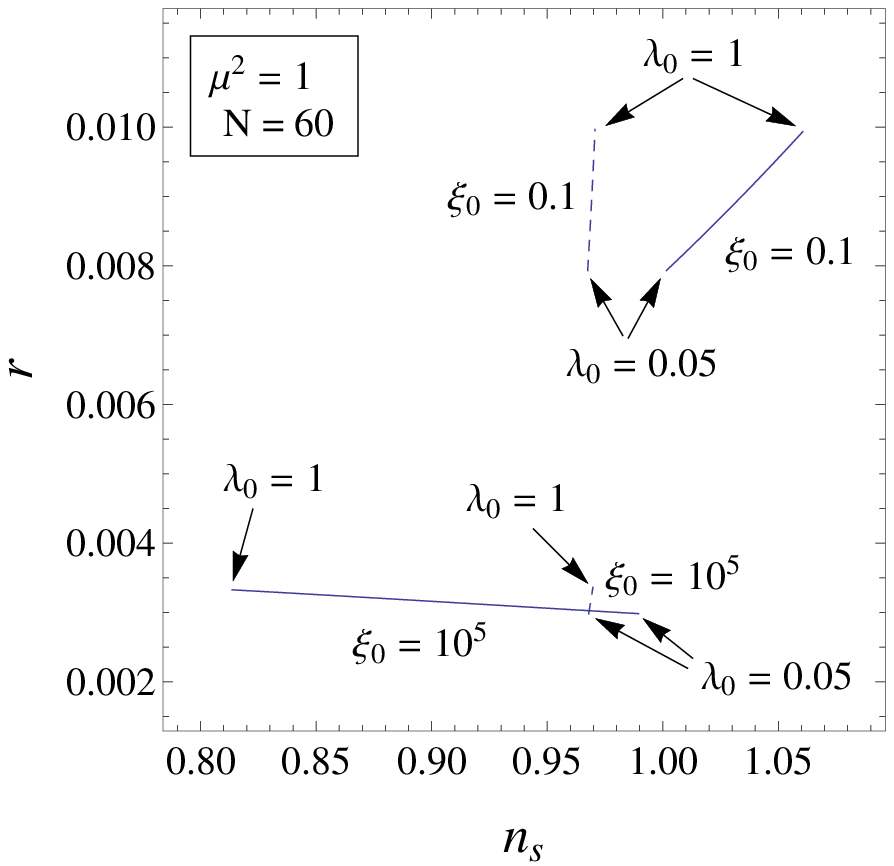}
  \end{center}
    \caption{Behavior of $n_s$ and $r$ as a function of the scalar quartic coupling, $\lambda_0$, in the interval from $0.05$ to $1$ for $N=60$, $\mu^2=1$ and $\xi_0 = 0.1,10^5$. The solid and dashed lines are drawn for the two-loop and leading-log results respectively.}
\label{fig:nsr:lambda0}
 \end{minipage}
  \hspace{0.04\hsize}
 \begin{minipage}{0.48\hsize}
  \begin{center}
   \includegraphics[width=71mm]{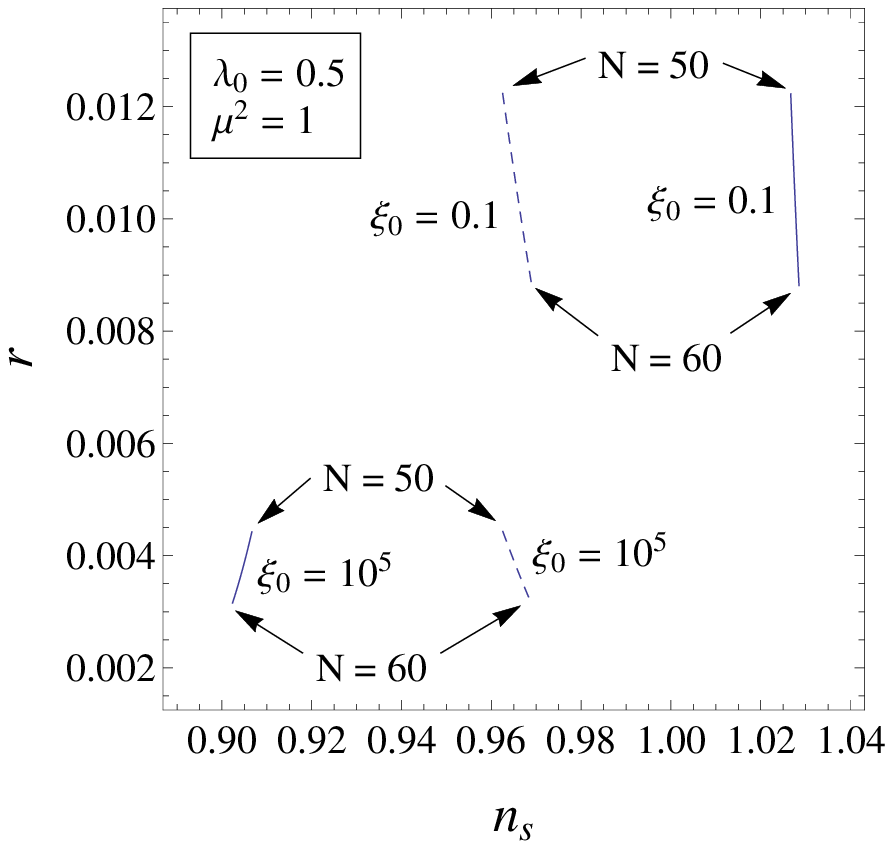}
  \end{center}
    \caption{Behavior of $n_s$ and $r$ as a function of the e-folding number, $N$, in the interval from $50$ to $60$ for $\mu^2=1$, $\lambda_0=0.5$ and $\xi_0 = 0.1,10^5$. The solid and dashed lines are drawn for the two-loop and leading-log results respectively.}
\label{fig:nsr:ef}
 \end{minipage}
\end{figure}

Fig. \ref{fig:nsr:lambda0} represents the $\lambda_0$-dependence of the inflationary parameters. Since the two-loop effective potential (\ref{pot}) is perturbatively calculated, we vary $\lambda_0$ in the interval from $0$ to $1$. It is found that the tensor-to-scalar ratio, $r$, increases with the growth of $\lambda_0$. The $\lambda_0$-dependence of $n_s$ is enhanced for a larger $\xi_0$ in the two-loop result, which is again in contrast to the leading-log result. 

We plot the e-folding number dependence in Fig. \ref{fig:nsr:ef}, shifting $N$ from $50$ to $60$, in which it is sufficient to solve the flatness or horizon problem. We see the tensor-to-scalar ratio, $r$, decreases as the e-folding number, $N$, increases. The range of $r$ is almost equivalent for both two-loop and leading-log results. Furthermore we verify the suppression of $N$-dependence for a larger $\xi_0$.

The inflationary parameters are constrained from the observations. Current observational constraints on the inflationary parameters are given in Table \ref{table:obs}. The scalar fluctuations and the spectral index have been observed by Planck and a recent joint analysis of BICEP2/Keck with Planck. Only an upper limit is obtained for the tensor-to-scalar ratio, $r$.
\begin{table} [h]
\begin{center}
\begin{tabular}{c|c|c|c}
 Observations & $\delta \times 10^5$ & $n_s$ & $r$ \\ \hline
 Planck + WP + lensing & 4.93 & $0.9653\pm0.0069$ & $<0.13$ \\
 BICEP2/Keck + Planck & --- & --- & $<0.12$ \\ 
\end{tabular}
 \caption{Constraints on inflationary parameters by Planck \cite{Ade:2013uln,Wang:2014kqa} and BICEP2/Keck with Planck \cite{Ade:2015tva}.}
 \label{table:obs}
\end{center}
\end{table}

We try to explain these constraints under the two loop effective potential. For the intervals of the four model parameters
\begin{equation}
0.005<\xi_0<10^5,\ 0\leq\lambda_0\leq1,\ 10^{-6}<\mu^2<10^3,\ 50\leq N\leq60,
\end{equation}
we search suitable parameter sets. As for the effective scalar-dependent quartic coupling we also keep
\begin{equation}
0\leq\lambda\leq1.
\end{equation}

It is possible to reproduce the values of the scalar fluctuations and spectral index obtained by Planck and WMAP. In Table~\ref{table:model} we summarize the parameter sets which are consistent with the observational data for the scalar fluctuations and the spectral index. The tensor-to-scalar ratio does not exceed the upper limit by BICEP2/Keck and Planck. It is found that the scalar-curvature coupling, $\xi_0$, and the renormalization scale, $\mu^2$, play a crucial role in determining the inflationary parameters. It is noted again that the case with $\mu^2=10^{-6}$ is appropriate for a typical energy scale of inflation, about $10^{15}$GeV. 
\begin{table} 
\begin{center}
\begin{tabular}{c|c|c|c|c}
 $(\xi_0,\lambda_0,\mu^2,N)$ & $\lambda$ & $\delta \times 10^5$ & $n_s$ & $r$ \\ \hline
 $(44.4,0.0383,10^{-6},60)$ & $0.03850$ & $4.9332$ & $0.96529$ & $0.002991$ \\
 $(44.4,0.0385,10^{-6},50)$ & $0.03870$ & $4.9282$ & $0.96531$ & $0.004227$ \\ \hline
 $(91.8,0.17,1,60)$              & $0.16998$ & $4.9306$ & $0.96534$ & $0.003038$ \\
 $(93.6,0.1775,1,50)$          & $0.17742$ & $4.9329$ & $0.96531$ & $0.004285$ \\ \hline
 $(150.6,0.5,13.9,60)$         & $0.49264$ & $4.9307$ & $0.96529$ & $0.003166$ \\
 $(151.5,0.5,11.65,50)$       & $0.49262$ & $4.9300$ & $0.96529$ & $0.004433$ \\ \hline
 $(200.6,1,27.38,60)$          & $0.96309$ & $4.9306$ & $0.96533$ & $0.003350$ \\
 $(203.6,1,22.88,50)$          & $0.96293$ & $4.9302$ & $0.96534$ & $0.004649$ \\
\end{tabular}
 \caption{Our results of inflationary parameters.}
 \label{table:model}
\end{center}
\end{table}

 \section{Conclusions}

In this paper we have studied a two-loop $\phi^4$ theory with non-minimal scalar-curvature coupling. Assuming the slow-roll scenario, we have analytically derived the expressions for the inflationary parameters, the scalar fluctuations, $\delta$, the spectral index, $n_s$, and the tensor-to-scalar ratio, $r$. The Lyaounov stability has been evaluated for a negative $\xi_0$. In this region the de Sitter solutions become unstable when the absolute values of the couplings decrease. 

The behavior for $n_s$ and $r$ has been systematically evaluated as a function of the couplings, the renormalization scale and the e-folding number. It is found that the two-loop corrections on $n_s$ are larger than the leading-log corrections. Besides, it is demonstrated that a larger $\lambda_0$- and $\mu^2$-dependence of $n_s$ is observed as the scalar-curvature coupling, $\xi_0$, increases. We have also observed a similar feature for $\xi_0$-, $\mu^2$- and $N$-dependence of $r$. Concretely the tensor-to-scalar ratio $r$ is suppressed with the growth of $\xi_0$, $\mu^2$ and $N$.

The loop corrections introduce an unavoidable dependence on the renormalization scale. It is observed that the two-loop contribution is softened for $\mu^2=\phi_N^2$. It should be noted that we can tune the couplings, $\lambda_0$ and $\xi_0$ consistent with the current observational constraints for $n_s$ and $r$.
In SM Higgs inflation the quartic coupling vanishes near the Planck scale \cite{Hamada:2014wna}. A small positive value of the non-minimal scalar curvature coupling is favored for a simple scalar quadratic potential \cite{Boubekeur:2015xza}.

It has been reported in Ref. \cite{Kallosh:2013tua} that there is an attractor in the $(n_s,r)$ plane at the large $\xi_0$ limit. We find that such attractor behavior does not appear in our model until $\xi_0=10^5$. But the results seem to approach the attractor.  There is a possibility that a larger $\xi_0$ is necessary to obviously observe the convergence to the attractor due to the quantum effects.
It has been pointed that multi-field models with quartic self-couplings quickly relax to the single-field attractor for a wide range of couplings and initial conditions \cite{Kaiser:2013sna}. Thus it is interesting to extend our approach to multi-field models.  

It is mentioned that the formulation of scalar-tensor theories and the theoretical predictions depend on the choice of the frames, Jordan or Einstein frame \cite{Faraoni:1999hp}. Note that calculation of inflationary parameters and Lyapunov stability may 
look to depend on the Einstein or Jordan frame under consideration. However, it has been proved in \cite{Kaiser:1994wj,Kaiser:1994vs} that the obtained inflationary parameters are just the same in both frames. 
In Ref. \cite{Kallosh:2014laa} the induced inflation models are revealed to have another attractor at the small $\xi_0$ limit. The radiative corrections may affect such a double attractor behavior. Note finally that using same approach one can study non-minimal two-loop 
inflation for Standard Model. This will be considered elsewhere.

\section*{Acknowledgements}
The authors would like to thank K.~Ishikawa and K.~Yamamoto for fruitful discussions. SDO would like to thank A.~Linde for clarifying remarks.The work by T.I. is supported in part by JSPS KAKENHI Grant Number 26400250 and that by SDO is supported in part by MINECO (Spain), projects FIS2010-15640 and FIS2013-44881 and by the  Program of Competitive Growth of Kazan Federal University.

\end{document}